\documentclass[
   aps,
   prb,
   preprint,
   groupedaddress,
   amsmath,amssymb
]{revtex4-2}

\usepackage{color}
\usepackage{graphicx}

\begin{document}

\title{Understanding the $sp$ magnetism in substitutional doped graphene}

\author{J. Hern\'andez-Tecorralco}
\email{juanht@ifuap.buap.mx}
\affiliation{Instituto de F\'isica, Benem\'erita Universidad Aut\'onoma de Puebla, Apartado Postal J-48, 72570, Puebla, Puebla, M\'exico
}%
\affiliation{Departamento de F\'isica Aplicada, Centro de Investigaci\'on y de Estudios Avanzados del IPN, Apartado Postal 73, Cordemex, 97310, M\'erida, Yucat\'an, M\'exico
}

\author{L. Meza-Montes}
\affiliation{Instituto de F\'isica, Benem\'erita Universidad Aut\'onoma de Puebla, Apartado Postal J-48, 72570, Puebla, Puebla, M\'exico
}
  
\author{M. E. Cifuentes-Quintal}
\affiliation{Departamento de F\'isica Aplicada, Centro de Investigaci\'on y de Estudios Avanzados del IPN, Apartado Postal 73, Cordemex, 97310, M\'erida, Yucat\'an, M\'exico
}

\author{R. de Coss}
\email{romeo.decoss@cinvestav.mx}
\affiliation{Departamento de F\'isica Aplicada, Centro de Investigaci\'on y de Estudios Avanzados del IPN, Apartado Postal 73, Cordemex, 97310, M\'erida, Yucat\'an, M\'exico
}
\affiliation{Facultad de Ciencias en F\'isica y Matem\'aticas, Universidad Aut\'onoma de Chiapas, 29050 Chiapas, M\'exico.}

\date{\today}

\begin{abstract}

Defect-induced magnetism in graphene has been predicted theoretically and observed experimentally. However, there are open questions about the origin of the magnetic behavior when substitutional impurities with $sp$ electrons are considered. The aim of this work is to contribute to the understanding of impurity-induced spin magnetism in doped graphene systems. Thus, the electronic structure and spin magnetic moments for substitutional doped graphene with impurities from groups IIIA (B, Al, and Ga) and VA (N, P, As, Sb, and Bi) of the periodic table were obtained within the framework of density functional theory. The nature of the magnetic ground state was determined from calculations of the total energy as a function of the spin magnetic moment using the fixed spin moment method. We show that the spontaneous magnetization in the studied systems arises from an electronic instability by the presence of a narrow impurity band at the Fermi level. Furthermore, we found that the emergence of spin polarization requires the impurity to introduce an extra electron to the graphene lattice and that the impurity-carbon hybridization is close to the $sp^3$ geometry. These features reveal that the charge doping sign and the hybridization degree play a fundamental role in the origin of $sp$ magnetism in substitutional doped graphene.

\end{abstract}

\maketitle

\section{Introduction}

Although pristine graphene is a non magnetic material \cite{novoselov}, nowadays there is a considerable amount of research about different mechanisms to induce magnetism in graphene-based systems \cite{yazyev, gonzales, tucek}. The magnetic properties exhibited in defective graphene are mainly the result of vacancies, impurities, and edge states. Particularly, the $sp$-electron magnetism has caused considerable interest because of two main reasons: (i) the electrons involved in the magnetic behavior are $s$ and $p$, rather than localized $d$ or $f$, and (ii) the possibility of room-temperature magnetism \cite{edwards}. Substitutional doping in graphene with $sp$ impurities is one route in the search for $sp$-electron magnetism. Substitutional doping consists of replacing one carbon atom with different single chemical species. Several theoretical studies based on density functional theory (DFT) calculations for single-substitutional $sp$ impurities in graphene have been performed (B, N, Al, Si, P, S, Ga, Ge, As, Se, Sb, Bi) \cite{
yzhou,pasti,n-graphene, dai-1,dai-2, denis-1, denis-2, sb-graphene, sb-graphene2, sb-graphene3, bi-graphene, wang, langer,jht}, showing that only a few of those impurities induce magnetism (P, N). The first case is P-graphene, which has spontaneous polarization with a spin magnetic moment of 1.0$\mu_ B$ per impurity \cite{dai-1,dai-2,denis-1,wang,jht}. Whereas the magnetic structure shows a local curvature around the P atom, a paramagnetic planar structure is found with higher energy than the distorted one \cite{dai-1}.
A second case is N-graphene which is particularly interesting because DFT calculations with the generalized gradient approximation (GGA) for the exchange-correlation functional show a paramagnetic character \cite{yzhou,pasti,dai-1}, while calculations using a meta-GGA functional reveal magnetism \cite{n-graphene}. This raises questions about whether meta-GGA functionals should be used to describe magnetic materials. Interestingly, it was recently shown that the use of meta-GGA does not improve the magnetic description of itinerant magnets over GGA functionals \cite{tran}.

On the other hand, two representative experimental reports have confirmed that doped graphene with impurities of P \cite{lin} and N \cite{miao} exhibits ferromagnetism. By thermal annealing of fluorographite in P vapor, samples of doped graphene were synthesized with high P concentration (2.86-6.40 at.\%) \cite{lin}. Magnetic measurements performed by means of a superconducting quantum interference device magnetometer evidenced a magnetic ordering driven by the presence of P groups. In the case of N impurities, Miao \textit{et al}. studied samples of doped graphene prepared by a self-propagating high-temperature synthesis at different concentrations (up to 11.17 at.\% of N) \cite{miao}. Their results revealed ferromagnetism with a Curie temperature greater than 673 K for samples with high N content. Nevertheless, these experimental samples contain different functional groups, which makes it difficult to know precisely the origin of the magnetic order. In addition, as far as we know, experimental evidence of magnetism in a single-substitutional impurity in graphene is still lacking.

It is important to note that the nature and origin of impurity-induced $sp$ magnetism in substitutional doped graphene have not been fully explained \cite{n-graphene, dai-1,dai-2,denis-1,wang,langer}. Well-known arguments based on Lieb's theorem \cite{lieb} about a sublattice imbalance between $\pi$ states are insufficient to explain why only some impurities could induce magnetism in graphene. To the best of our knowledge, there are no further reports about spontaneous magnetization in other substitutional doped graphene structures with $sp$ impurities like for P and N \cite{yzhou, pasti, denis-2, sb-graphene, sb-graphene2, sb-graphene3, bi-graphene}. Therefore, the aim of this work is to contribute to the further understanding of impurity-induced $sp$ magnetism in substitutional doped graphene. In this work, we present results based on first-principles calculations for the electronic and magnetic properties of different doped graphene systems with B, Al, Ga, N, P, As, Sb, and Bi as impurity atoms. First, from the analysis of the electronic structure in the paramagnetic state, we notice the emergence of an impurity band at the Fermi level and the role of its bandwidth in the rise of spontaneous magnetization. A narrow impurity band causes an electronic instability which favors a magnetic state, but the spontaneous magnetization requires the impurity to introduce an extra electron in the graphitic lattice. Second, we analyze the role of the charge-doping sign and the  impurity-carbon hybridization in the impurity bandwidth and, consequently, in the emergence of magnetism in substitutional doped graphene. This paper is organized as follows: Sec. \ref{compdetails} gives a description of the computational details applied in our work. The results are presented and discussed in Sec. \ref{results}, which is divided into two parts; the first one presents the electronic and magnetic properties of substitutional doped graphene. In the second part, we examine and discuss the role of the impurity-carbon hybridization geometry in the emergence of magnetism. Finally, our main findings are summarized in Sec. \ref{summary}.

\section{Computational Details}
\label{compdetails}

Total energies, electronic structure and spin magnetic moments were determined by solving self-consistently the Kohn-Sham equations within the framework of plane waves and pseudopotentials as implemented in the \texttt{QUANTUM ESPRESSO} code  \cite{QE1,QE2}. Core electrons were replaced by ultrasoft pseudopotentials from the \texttt{GBRV} library \cite{garrity}, which has been optimized for use in high-throughput DFT calculations. Valence electron states were expanded in plane waves with a kinetic energy cutoff of 40 Ry and charge density cutoff of 320 Ry. The exchange-correlation functional was treated within the Perdew-Burke-Ernzerhof \cite{pbe} parametrization of the GGA.  In order to simulate an isolated layer, we left at least 15 \AA\ of vacuum space between periodic images. We simulated substitutional impurities by replacing one carbon atom of the graphene pristine lattice. A supercell of 8 $\times$ 8 unit cells was used in order to simulate an impurity concentration in the dilute limit  ($c < $ 1.0 at. \% ). We fixed the supercell lattice constant of each doped system at the corresponding ground-state lattice constant of pristine graphene. During all the structural calculations the atomic positions were relaxed with a Broyden-Fletcher-Goldfarb-Shanno quasi-Newton algorithm until the internal forces were less than 0.01 eV/\AA. For structural optimization of the supercells, the Brillouin zone was integrated with a Monkhorst-Pack mesh \cite{kpoints} of 3 $\times$ 3 $k$ points with a Methfessel-Paxton \cite{integration} smearing of 0.01 Ry. However, for the calculation of the electronic structure and the spin magnetic moments we used a 9 $\times$ 9 mesh with a smearing of 0.002 Ry to describe accurately the electronic and magnetic properties. The effect of spin-orbit coupling (SOC) on the electronic and magnetic properties was analyzed for graphene with substitutional N, P, As, Sb, and Bi, which showed no significant difference in the bands close to the Fermi level and their magnetic moments with respect to the case without SOC (see the Supplemental Material \cite{sm}\nocite{pseudo2}). Henceforth, we present only results neglecting the SOC for all the studied systems.

\section{Results and Discussion}
\label{results}

\subsection{Impurity-induced magnetism in substitutional doped graphene}

\begin{figure}[!ht]\centering
	\includegraphics[scale=1.0]{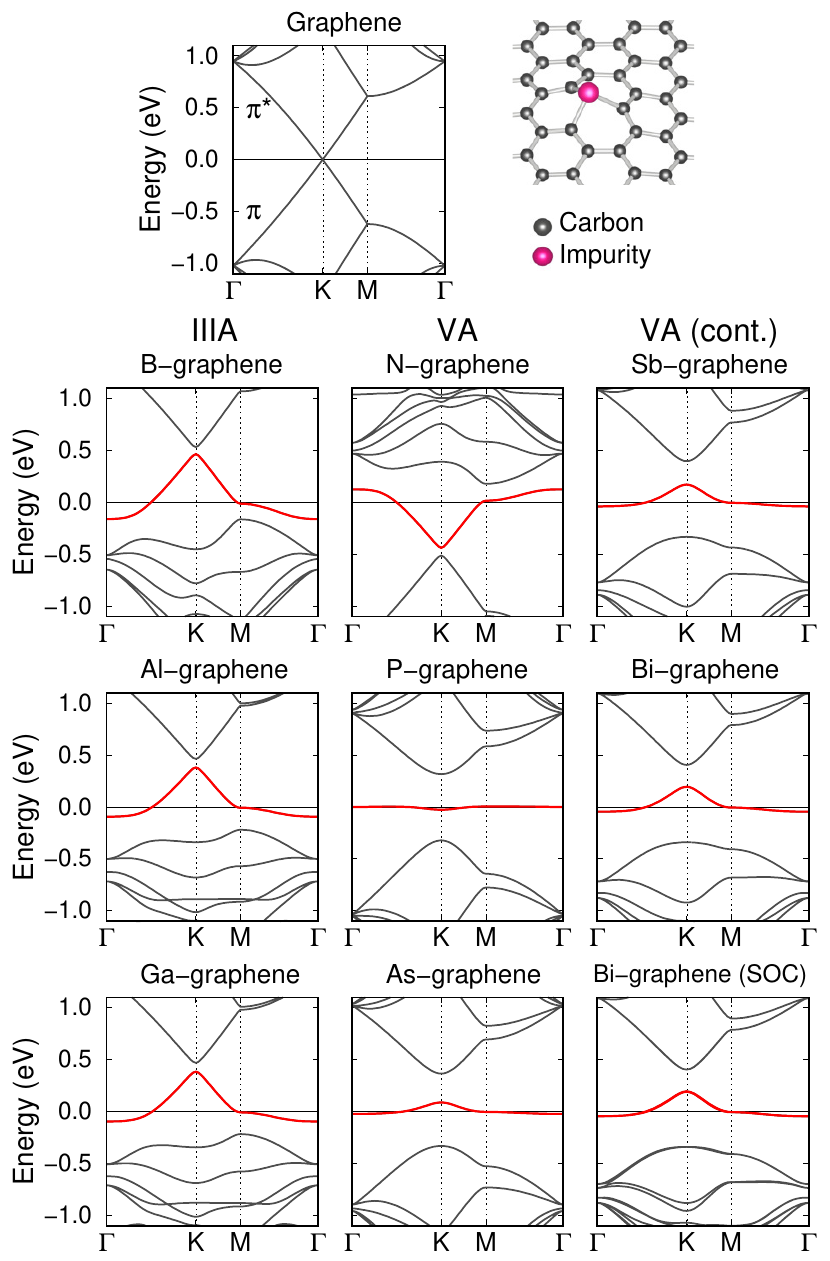}
	\caption{Paramagnetic band structures of doped graphene with \textit{sp} impurities from groups IIIA (B, Al, Ga) and VA (N, P, As, Sb, Bi). The band structure of pristine graphene is shown at the top. The origin of the energy scale has been set at the Fermi level $E_F$. The impurity band state is highlighted in red. Bi-graphene considering spin-orbit coupling (SOC) is in the bottom right.}
	\label{Fig1}
\end{figure}

In order to examine the effect of different $sp$ impurities on the magnetic behavior of doped graphene, we performed DFT calculations of single substitutional doped graphene systems with impurities from groups IIIA (B, Al, Ga) and VA (N, P, AS, Sb, Bi) of the periodic table. Figure \ref{Fig1} shows the paramagnetic band structures of pristine and substitutional doped graphene. At the top of Fig. \ref{Fig1}, the folded band structure of pristine graphene shows the characteristic Dirac cone around the Fermi level formed in the \textit{K} point by the well-known $\pi$ and $\pi^*$ bands. With doping, the Fermi level is moved to the occupied $\pi$ band for the hole-doped case  (B, Al, Ga) or the unoccupied $\pi^*$ band for the electron-doped case (N, P, As, Sb, Bi). The formation of an impurity band at the Fermi level induced by the substitutional doping is highlighted (band in red in Fig. \ref{Fig1}). From the band structure, in the hole-doped cases, the impurity band has a similar behavior regardless of the impurity, whereas, for the electron-doped case, the impurity band depends on the specific impurity. This impurity band can be characterized by its bandwidth dispersion $W_{imp}$, and an examination of the bandwidth for all cases reveals that the electron-doped systems, except for N-graphene, have the smallest bandwidths and a trend to increase with the atomic number (see Table \ref{table1}). These results underline an electron-hole asymmetry in the electronic structure from doping graphene. 

\begin{table}
	\caption{\label{table1} Bandwidth dispersion of the impurity band $W_{imp}$, estimated Stoner parameter $I_S$, reformulated Stoner condition $I_S N_0/W_{imp}$, impurity height $h$ with respect to the first carbon neighbors,  hybridization angle $\theta$, and spin magnetic moment $M$ for each substitutional doped graphene system.}
\begin{ruledtabular}
\begin{tabular}{llllllll}
Impurity & $W_{imp}$ (eV) & $I_S$ (eV) & $I_S N_0 /W_{imp}$ & $h$ (\AA) & $\theta$ (deg) & $M$ (units of $\mu_B$/cell) \\
\hline
B & 0.626 & 0.535 & 0.855  & 0.00 & 120.0 & 0.0  \\
 Al & 0.474 & 0.427 & 0.901 & 0.77 & 103.8 & 0.0  \\
  Ga & 0.480 & 0.428 & 0.892 & 0.72 & 105.5 & 0.0 \\
   N  & 0.563 & 0.486 & 0.863 & 0.00 & 120.0 & 0.0 \\
		P  & 0.032 & 0.090 & 2.812 & 0.80 & 100.7 & 1.0 \\
		 As & 0.111 & 0.147 & 1.324 & 1.05 & 92.8 & 1.0  \\
			Sb & 0.210 & 0.234 & 1.114 & 1.32 & 84.7 & 0.7 \\
			 Bi & 0.240 & 0.259 & 1.079 & 1.46 & 80.7 & 0.6 \\
\end{tabular}
\end{ruledtabular}
\end{table}

\begin{figure}[!ht]\centering
	\includegraphics[scale=1.0]{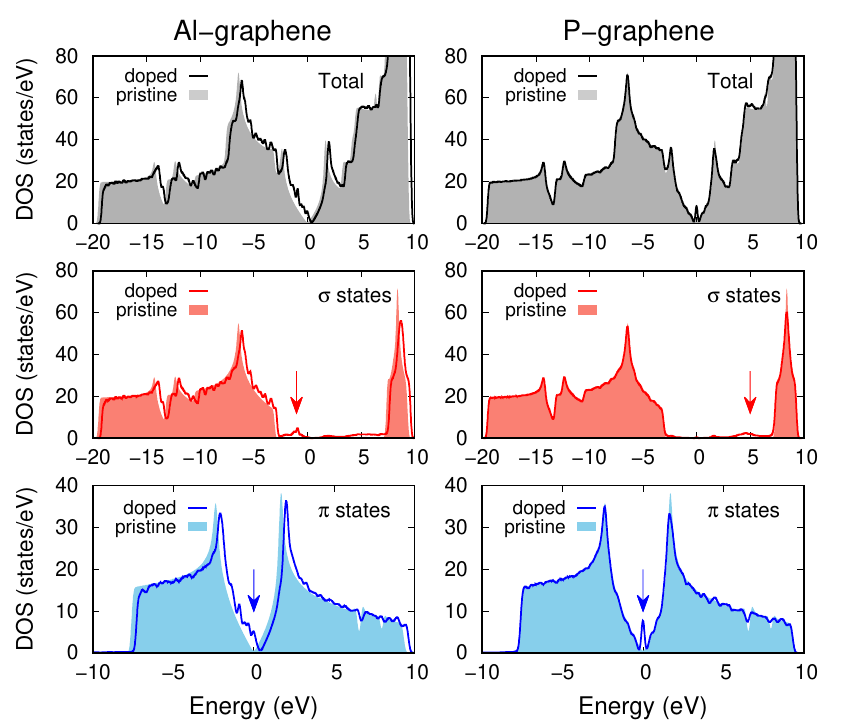}
	\caption{Total and projected (by $\sigma$ and $\pi$ bands) densities of states (DOS) for Al-graphene and P-graphene. Solid lines correspond to the doped system. For comparison, the shaded area corresponds to the DOS of pristine graphene. Arrows indicate the new states in the doped case. Note that the panels at the bottom have different scales. }
	\label{Fig2}
\end{figure}

To have better comprehension of the electron-hole asymmetry in the electronic structure in doped graphene, we calculated the paramagnetic density of states (DOS) of Al-graphene and P-graphene. Figure \ref{Fig2} shows the total DOS as black solid lines, whereas the projected DOSs with $\sigma$ and $\pi$ character are shown by red and blue solid lines, respectively. The $\sigma$ states are a result of the hybridization of $s$, $p_x$, and $p_y$ orbitals, whereas the $\pi$ states are delocalized states formed by $p_z$ orbitals. For comparison, the shaded areas in Figure \ref{Fig2} correspond to the DOS of pristine graphene. The total density of states shows significant differences between the hole-doped and electron-doped cases. For Al-graphene, the electronic states suffer an energy shift with respect to the electronic bands of pristine graphene as expected for hole doping. In contrast, P-graphene shows the emergence of a prominent peak at the Fermi level. An analysis of the projected  $\sigma$ bands of the DOS reveals that the impurity induces new states for the hole-doped and electron-doped cases. For Al-graphene these states are mainly localized just below the Fermi level in the range of -4.0 to -2.0 eV, whereas for P-graphene, these states are found in the range of 3.0 to 6.5 eV in the unoccupied bands. Closer inspection of the projected $\pi$ bands of the DOS shows that the sharp peak at the Fermi level in P-graphene comes from these states. Al-graphene also presents $\pi$ states at the Fermi level. The main difference is that the sharp peak in the electron-doped case is narrower than in the hole-doped case, as observed in the band structures. 

\begin{figure}[!ht]\centering
	\includegraphics[scale=1.0]{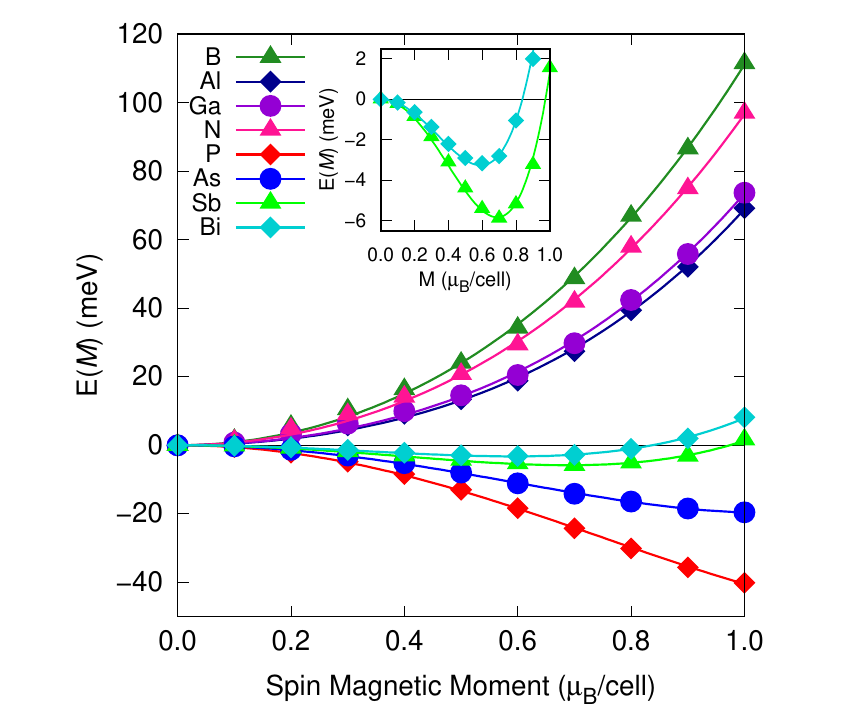}
	\caption{Energy as a function of the magnetic moment for  doped graphene with B, Al, Ga, N, P, As, Sb, and Bi, calculated using the fixed spin moment method. The solid lines correspond to the fourth-order polynomial fit of even powers of $E(M)$ to the calculated values (symbols).}
	\label{Fig3}
\end{figure}

In order to investigate the magnetic ground state of each substitutional doped structure, we have carried out calculations of total energy as a function of the spin magnetic moment $M$ using the fixed spin moment (FSM) method \cite{mohn-fsm}. Figure \ref{Fig3} shows the calculated values of the total energy $E$ (symbols) as a function of $M$. The total energy $E(M)$ is referred to the paramagnetic state ($M=0.0$), and the solid lines are a fourth-order polynomial fit of even powers as in Landau's free-energy expression, $E(M) = E_0 +  \alpha M^2 + \beta M^4$ \cite{moruzzi}. The inset is a close-up of the $E(M)$ curves for Sb- and Bi-graphene. As  illustrated in Fig. \ref{Fig3},  the three hole-doped systems and N-graphene have a minimum energy at $M= 0.0$, indicating a paramagnetic ground state, whereas four electron-doped systems have a minimum energy value for $M > 0$. The FSM reveals that for the electron-doped systems, with the exception of N-graphene, the ground state is magnetic. Self-consistent spin-polarized calculations and the FSM curves indicate that P-graphene and As-graphene have an integer magnetic moment of $M=$ 1.0$mu_B$/cell [for $M > 1.0$, $E(M)$ increases  abruptly]; meanwhile, Sb-graphene and Bi-graphene have fractional magnetic moments of $M$ = 0.7$\mu_B$/cell and $M=$ 0.6$\mu_B$/cell, respectively. These results are summarized in Table \ref{table1}. It is important to underline that, aside from the observed magnetism in P-graphene, we found that heavier elements from the VA group such as As, Sb, and Bi also induce a net magnetic moment in graphene. 

\begin{figure}[!ht]\centering
	\includegraphics[scale=1.0]{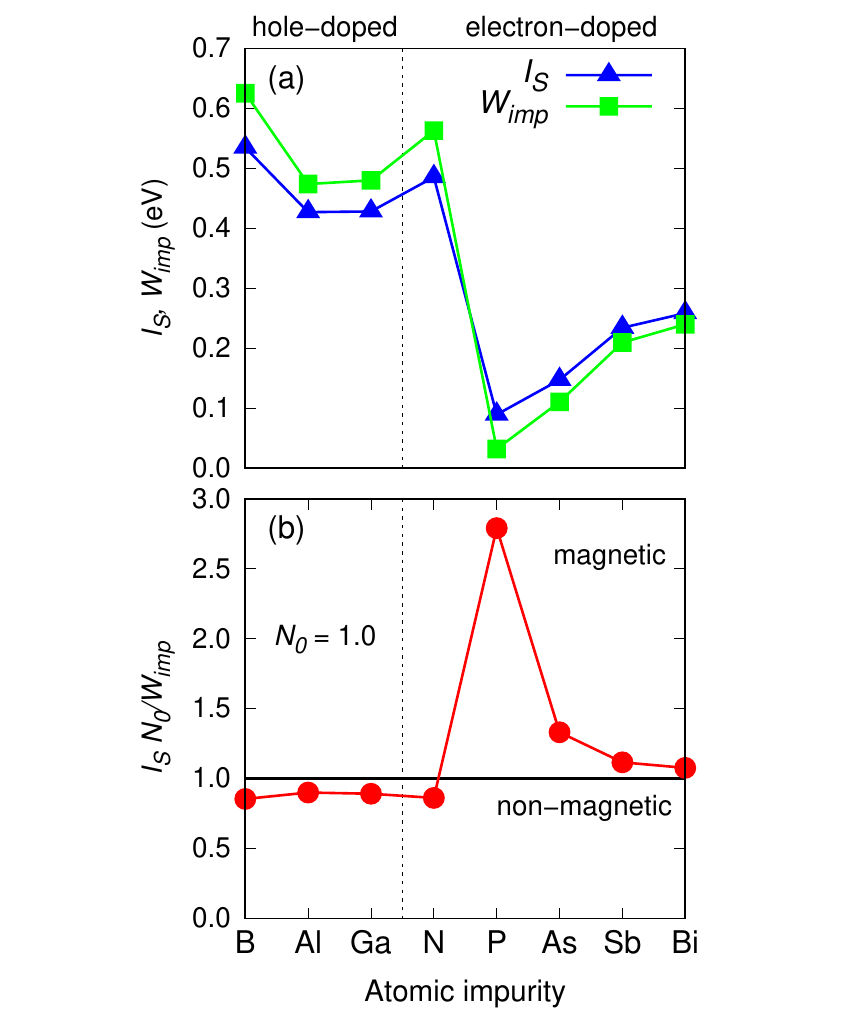}
	\caption{ (a) The estimated value of the Stoner parameter $I_S$ and the calculated impurity bandwidth $W_{imp}$ for each impurity. (b) $I_S N_0/W_{imp}$ ratio for each doped graphene system. }
	\label{Fig4}
\end{figure}

The study of $sp$ magnetism in substitutional doped graphene has demonstrated the correlation between the impurity bandwidth and the origin of spontaneous magnetization, where small values of $W_{imp}$ favor the formation of a magnetic moment \cite{jht}. In the group of impurities under study, P-graphene has the smallest bandwidth (32 meV), thus showing magnetism, as previously reported \cite{dai-1,dai-2,denis-1, wang,jht}. Therefore, we can anticipate the turning of magnetism in graphene with impurities that yield narrow bandwidths. Thus, to address the question of why only some $sp$ impurities induce a spontaneous magnetization in graphene, we return to the analysis of the impurity band. For all the systems under study, the integration of  the paramagnetic DOS up to the Fermi level of the impurity band reveals an occupancy of $N_0$ = 1.0 electrons, indicating a half-filling band. According to the Stoner model of ferromagnetism, a high value of the paramagnetic DOS at the Fermi level $N(E_F)$ makes a paramagnetic state unstable with respect to the magnetic phase if the Stoner criterion is satisfied, $I_S N(E_F) > 1$, where $I_S$ is the Stoner parameter \cite{moriya-book}. A quantitative estimation of $I_S$ can be obtained from fitting the fourth-order polynomial energy expression in the FSM curves previously described. The coefficient of the quadratic term  $\alpha$ is related to the Stoner parameter as $I_S = 1/N(E_F) - \alpha$ \cite{moriya-book}. For our analysis, we can approximate the impurity band as a narrow rectangular band \cite{jht}. This approach allows writing the paramagnetic DOS at the Fermi level as $N(E_F) = N_0/W_{imp}$  \cite{jht, edwards, gruber}, and consequently, we obtain an alternative expression for the Stoner criterion such as  $I_S N_0/W_{imp} > 1$. This expression allows us to rationalize the origin of the magnetism in substitutional doped graphene with a Stoner-type condition in terms of the impurity bandwidth $W_{imp}$. Figure \ref{Fig4}(a) shows the estimated value of the Stoner parameter $I_S$ in a rectangular band approximation and the impurity bandwidth $W_{imp}$ obtained from the paramagnetic band structure. Except for the anomalous case of N-graphene, two behaviors are clearly distinguished for the hole-doped and electron-doped systems. Interestingly, under this approach, the parameter $I_S$ seems to be dependent on the type of impurity just like $W_{imp}$. Figure \ref{Fig4}(b) shows the result of applying the reformulated Stoner criterion for all studied cases in a half-filling band ($N_0=1.0$). Whereas the hole-doped systems and N-graphene do not fulfill the Stoner condition of magnetism, the rest of the electron-doped cases satisfy the relationship $I_S/W_{imp} > 1$. The numerical data for the estimated Stoner parameter and the Stoner-type condition are presented in Table \ref{table1}.  From this quantitative analysis, it is clear that the spontaneous polarization in substitutional doped graphene is driven by small values of $W_{imp}$ as long as $W_{imp} < I_S$. The magnetism induced through an electronic instability by the presence of a sharp singularity at, or close to, the Fermi level is a well-known effect which has been observed in graphene \cite{yazyev, gonzales} with adsorbed hydrogen as well as monolayers of GaSe \cite{cao}, $\alpha$-SnO \cite{seixas},  InP$_3$ \cite{nmiao1}, and In$_2$Ge$_2$Te$_6$ \cite{nmiao2} by hole doping.

\begin{figure}[!ht]\centering
	\includegraphics[scale=1.0]{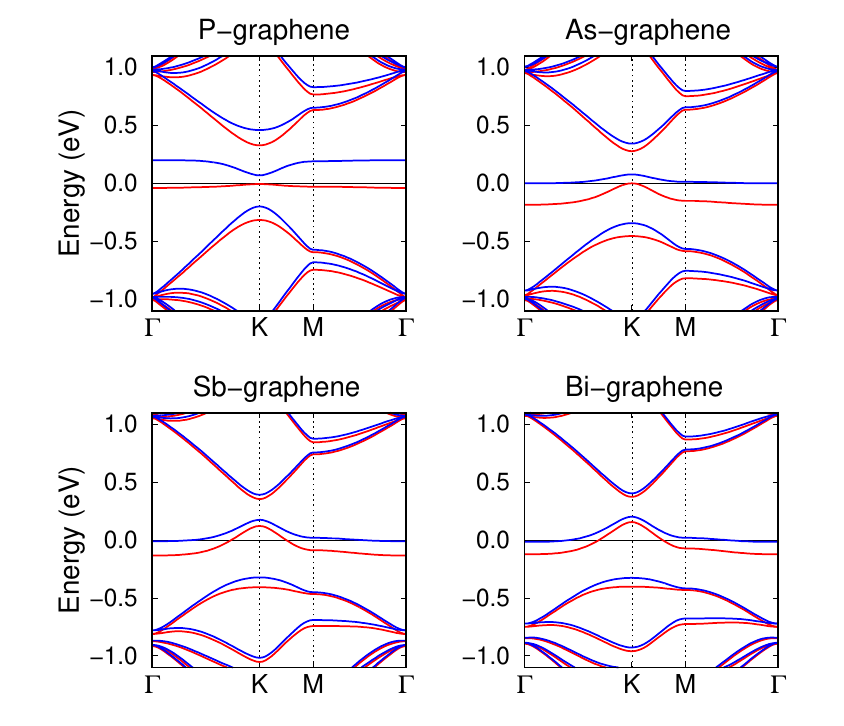}
	\caption{Spin polarized electronic band structure for substitutional doped graphene with P, As, Sb, and Bi. The red and blue lines correspond to  spin up and spin down, respectively. The origin of the energy scale has been set at the Fermi level $E_F$.}
	\label{Fig5}
\end{figure}

The electronic band structures for the spin-polarized cases are shown in Fig. \ref{Fig5}.  The spin-up states are in red, whereas the spin-down states are in blue. For P-graphene and As-graphene, we can see fully spin polarized bands since the spin-up band is fully occupied with one electron, whereas the spin-down band is empty. On the other hand, the spin splitting observed in the band structure of Sb-graphene and Bi-graphene shows partially occupied spin-up and spin-down bands. The bandwidth dispersion of the impurity band is different on each spin channel.  This difference evidences that the spin polarization of the impurity band is not a rigid band splitting. Our results show that the magnetic ground states of P-graphene and As-graphene are systems that exhibit a band gap opening in both spin channels, while Sb-graphene and Bi-graphene present a metallic character. From the spin-polarized bands, we can estimate the exchange splitting as  $\Delta_s = \sum_{FBZ} w_k|(\epsilon_{imp,k}^{\uparrow}-\epsilon_{imp,k}^\downarrow)|$, where $\epsilon_{imp}^{\uparrow}$ and $\epsilon_{imp}^\downarrow$, are the impurity band energies per spin channel and $w_k$ is their weight on each $k$ point in the  first Brillouin zone. The exchange splittings for substitutional doped graphene with P, As, Sb, and Bi are 0.199, 0.157, 0.101, and 0.087 eV, respectively. Comparing these values with the bandwidth dispersion of the impurity band in Table \ref{table1} for the electron-doped systems, we found that systems which have an impurity bandwidth smaller than the exchange splitting show an integer magnetic moment and those systems which have an impurity bandwidth larger than the exchange splitting have a fractional magnetic moment. Thus, the magnetic moment that exhibits each magnetic system is a result of the relation between $W_{imp}$ and $\Delta_s$, so that  we have integer values as long as $W_{imp}$ $<$ $\Delta_s$ and fractional values for $W_{imp}$ $>$ $\Delta_s$. 

The present results indicate that, from the group of substitutional impurities analyzed for graphene, a spontaneous magnetization requires that (i) the impurity must have an atomic configuration that allows introducing an extra electron to the graphene lattice, like the case of impurities in the VA column, and (ii) the interaction of the impurity on graphene should induce the formation of a narrow impurity band at the Fermi level. 

\subsection{The role of the impurity-carbon hybridization}

 In this section, we explore the role of the structural and bonding features in the magnetic behavior in doped graphene. To have more physical insights into the origin of a narrow impurity band that gives rise to spontaneous magnetization, we take as case of studies B-graphene and N-graphene. This choice follows mainly from the fact that N-graphene is an anomalous system compared with the electron-doped cases since it is non magnetic. Both systems are planar materials with a non magnetic phase. In contrast, the magnetic systems have a trigonal pyramid-type geometry with a bond angle less than 109.5$^\circ$  (see Table \ref{table1}). This obeys an impurity atomic size larger than the carbon atom, which causes significant distortions in the lattice structure. Meanwhile, the similar atomic radii of B and N compared with the carbon atom keep the planar structure of graphene. These structural differences can be interpreted as different types of hybridization between the impurity and the carbon atoms in graphene, where the B-C and N-C bonds present an $sp^2$ hybridization, and in the other cases, the impurity-carbon bond corresponds to an $sp^3$-type hybridization. 

\begin{figure}[!ht]\centering
	\includegraphics[scale=1.0]{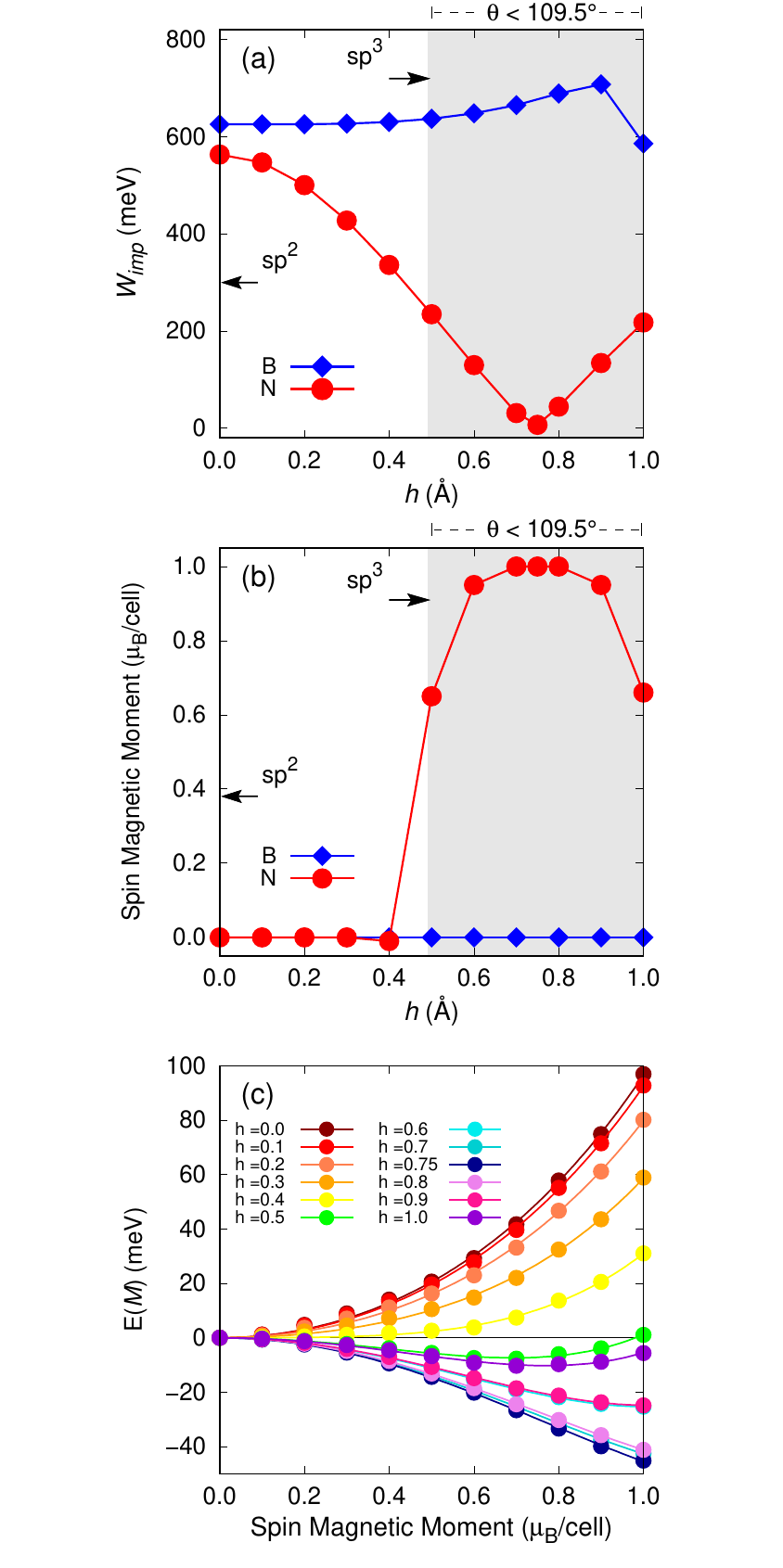}
	\caption{ (a) Evolution of the impurity bandwidth $W_{imp}$ and (b) the magnetic moment $M$ as a function of the impurity position $h$ for B-graphene and N-graphene. The magnetic regime is indicated in gray. (c) Energy as a function of the magnetic moment for different $h$ values calculated using the FSM method (symbols). The solid lines correspond to fourth-order polynomial fit of even powers of $E(M)$ to the calculated values.}
	\label{Fig6}
\end{figure}

With the aim of analyzing the structural effect of the impurity-carbon bond, we conducted simulations of B-graphene and N-graphene systems with different impurity-carbon hybridizations by artificially changing the position of the impurity atom with respect to planar graphene (height $h$). Thus, by fixing different values for $h$ we can analyze the transition from an $sp^2$ ($ h $ = 0, $\theta =120^\circ$) to $sp^3$ ($h > 0.5$ \AA\,, $\theta < 109.5^\circ$) hybridization.  As was discussed above, the bandwidth dispersion $W_{imp}$ of the impurity band plays a fundamental role in the origin of spontaneous magnetization. Therefore, $W_{imp}$ is a useful parameter for analyzing the evolution of the magnetism. Figure \ref{Fig6}(a) presents the values of $ W_{imp}$ obtained for different $h$ values from 0 to 1.0 \AA. The evolution of $W_{imp}$ reveals different trends for B-graphene and N-graphene. B-graphene has a slight change in its bandwidth, whereas N-graphene shows a significant decrease followed by an increase of $W_{imp}$ with a critical point at  0.75 \AA\, where the smallest bandwidth is found. These results anticipate that in B-graphene, regardless of the impurity-carbon hybridization geometry, the system remains paramagnetic. In contrast, we found that N-graphene could exhibit magnetism as along as $W_{imp}$ is small enough. To confirm these expectations, spin-polarized calculations were performed for both systems. Figure \ref{Fig6}(b) presents the spin magnetic moment as a function of $h$.  Accordingly, the spin-polarized calculations show that B-graphene is non magnetic regardless of the hybridization geometry. Instead, N-graphene shows a net magnetic moment for $h > 0.5$ \AA, i.e., when the impurity-carbon bond has an $sp^3$-type bond angle. According to these results, we can examine four regimes, one zone where the magnetic moment is zero ($M = 0$), followed by a second zone where the magnetic moment begins to increase with a fractional magnetic moment ($0 < M< 1.0$), a third zone where the magnetic moment is constant ($M=1.0$), and, finally, a zone where the magnetic moment begins to decrease with a fractional magnetic moment ($0 < M < 1.0$). The zones with a fractional magnetic moment correspond to the cases of partially occupied spin-up and spin-down bands, whereas the zone with $M=1.0\mu_B$/cell corresponds to the case of fully polarized bands. The FSM calculations of the total energy as a function of the spin magnetic moment in N-graphene for different $h$ are shown in Fig. \ref{Fig6}(c). The present results demonstrate the role of the impurity-carbon bond hybridization, where a system with an angle $\theta  < 109.5^{\circ}$ favors a small bandwidth dispersion of the impurity band. This affirmation is valid for only electron-doped cases. As we can see for B-graphene, in the hole-doped cases the hybridization is not related to spin polarization. This confirms the condition for magnetism in substitutional doped graphene; the impurity must introduce an extra electron. Last, the trend in $W_{imp}$ and $M$ described in the electron-doped systems can easily be extrapolated to the results presented in  Fig. \ref{Fig6} for N-graphene. The increase in atomic number in the VA column means an increase in the atomic radii,  which leads to different impurity heights and their corresponding impurity-carbon hybridization angles and, consequently, changes in $W_{imp}$ and $M$. As the electron-doped systems have impurity heights and impurity-carbon angles that correspond to cases with $sp^3$ hybridization, they are magnetic (see Table \ref{table1}).

\begin{figure}[!ht]\centering
	\includegraphics[scale=1.0]{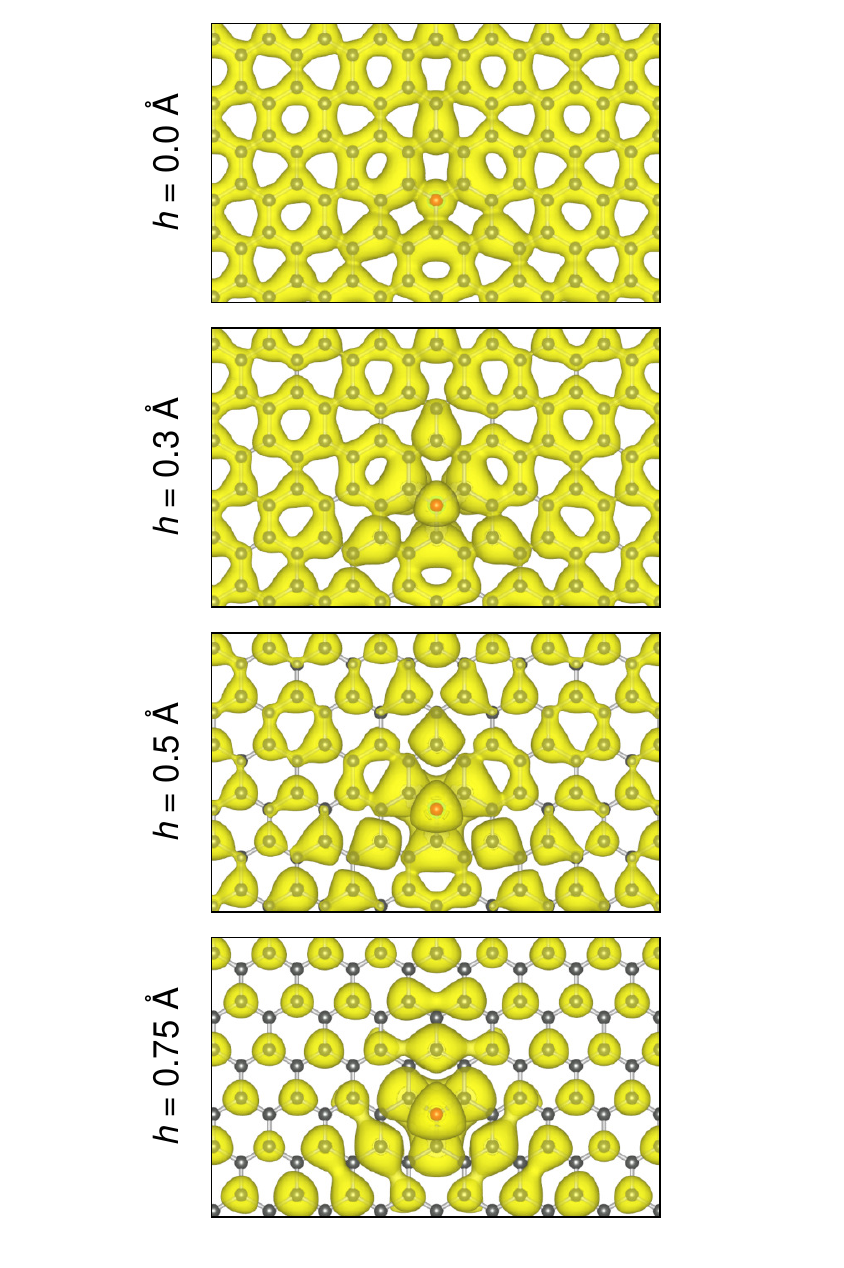}
	\caption{ Isosurface plots for the contribution of the impurity band to the charge density for N-graphene at four different $h$ (0.0, 0.3, 0.5, and 0.75 \AA). A lattice symmetry breaking of the electron charge density over sublattices is observed in $h$ = 0.75 \AA. The isosurface is shown at $10^{-3}$ \textit{e}/\AA$^3$.  }
	\label{Fig7}
\end{figure}

In an attempt to understand the origin of the impurity bandwidth behavior for different $h$ values, which gives rise to the magnetic character in N-graphene, we analyze the evolution of the impurity band wave function through electron charge density for B-graphene and N-graphene. The analysis of the impurity band charge density in the planar structures ($h = 0.0$) reveals the classic features of $\pi$ bands. The electron charge density is being distributed over the whole lattice suggests that the electron in the impurity band is spatially delocalized. The evolution of the impurity band charge density as a function of $h$ evidences two different mechanisms for the hole-doped and electron-doped cases. For B-graphene, when $h$ is increasing, the density around the impurity changes slightly, but in N-graphene the electron charge density exhibits a lattice symmetry breaking over sublattice A or B (see Fig. \ref{Fig7}). As we see in Fig. \ref{Fig6}(b) for N-graphene, for $h = 0.5$ \AA\ the system becomes magnetic. In Fig. \ref{Fig7}, it is interesting to note that for this $h$ value, the charge density distribution shows a clear difference over the two  different sublattices. Thus, we found that for N-graphene at  $h=0.5$ \AA\ the electron charge density is mainly concentrated in one sublattice, whereas in the adjacent sublattice it is reduced. In the critical point where the spin magnetic moment is maximum for N-graphene at $h$ = 0.75 \AA, the charge density is concentrated only on the impurity and the carbon atoms which belong to the sublattice adjacent to the impurity. This corresponds to a sublattice imbalance of the $\pi$ states. Since this effect does not occur in B-graphene, we can see why the impurity bandwidth is almost insensitive to the change in the hybridization geometry. These features show again the electron-hole asymmetry but in the impurity band charge density as a function of different impurity-carbon bond hybridizations. 

\begin{figure}[!ht]\centering
	\includegraphics[scale=1.0]{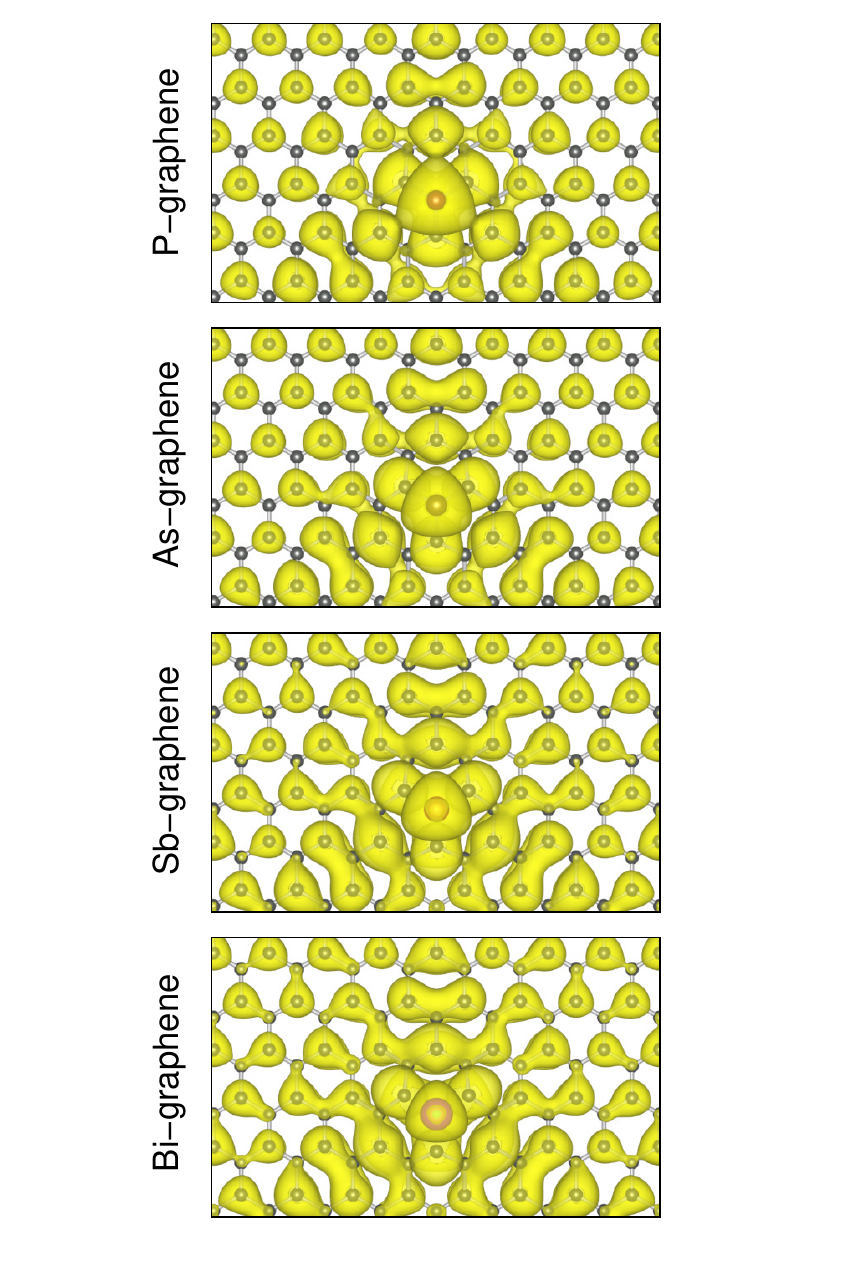}
	\caption{ Isosurface plots for the contribution of the impurity band to the charge density for P-graphene, As-graphene, Sb-graphene, and Bi-graphene. The isosurface is shown at $10^{-3}$ \textit{e}/\AA$^3$.  }
	\label{Fig8}
\end{figure}

To conclude, the impurity band charge densities for P-graphene, As-graphene, Sb-graphene, and Bi-graphene are shown in Fig. \ref{Fig8}. As previously discussed, P-graphene and As-graphene are systems with fully spin polarized bands, whereas Sb-graphene and Bi-graphene have partially spin polarized bands in the magnetic state. These features are related to the imbalance degree between $\pi$ states in the impurity band charge density, where for the first case there is a full imbalance over each sublattice and for the second one the imbalance of the $\pi$ states is partial. From these results, we show that for the electron-doped systems, the magnetism is associated with a sublattice imbalance of the $\pi$ states.  

\section{Summary}
\label{summary}

The analysis of paramagnetic band structures showed the appearance of an impurity band around the Fermi level in all the studied cases of doped graphene. For the hole-doped systems (B, Al, and Ga) the value of $W_{imp}$ is larger than 470 meV, while the electron-doped systems (N, P, As, Sb, and Bi) have a narrow impurity band with $W_{imp}  <$ 250 meV. From the analysis of the electronic structure for the doped systems we found that the hole-electron asymmetry in $W_{imp}$ is related to the fact that, in the hole-doped systems the impurity mainly introduces $\sigma$ states located below the Fermi level. In contrast, in the electron-doped systems, the impurity mainly introduces $\pi$ states, which are located at the Fermi level. With respect to the characterization of the magnetic states, the FSM curves showed that all the hole-doped cases and N-graphene have a non magnetic behavior, whereas the rest of the electron-doped systems have a net magnetic moment. The magnetic band structures for graphene doped with P and As show fully spin polarized bands, whereas those of Sb and Bi exhibit a partial spin polarization. Interestingly, we found that magnetic behavior is related to the sublattice imbalance degree of the $\pi$ states, where P-graphene and As-graphene exhibit full sublattice imbalance, while Sb-graphene and Bi-graphene exhibit only a partial one.

From the analysis of the electronic structure for N-graphene at different hybridization geometries, we showed that the electron-doped systems are magnetic as long as they present an $sp^3$-type hybridization. The analysis of the impurity band charge density in N-graphene revealed that the different impurity-carbon bond geometries are related to a sublattice imbalance of $\pi$ states in the emergence of spin polarization. Thus, the impurity-carbon hybridization is closely related to the emergence of magnetism in the electron-doped systems.

Finally, using a reformulated Stoner condition of magnetism for a narrow band, we showed that the spontaneous magnetization is driven by an electronic instability associated with a narrow impurity band at the Fermi level. Thus, we found that a narrow impurity band ($W_{imp} <$ 250 meV) is required in order to obtain spin polarization in doped-graphene systems. We note that this feature is present in the studied electron doped systems with an impurity-carbon hybridization close to the $sp^3$ geometry. From the present analysis, considerable insight has been gained with regard to the origin of the impurity-induced $sp$ magnetism in substitutional doped graphene.

\begin{acknowledgments}

This research was supported by Consejo Nacional de Ciencia y Tecnolog\'ia (Conacyt-M\'exico) under grant No. 288344. The authors thankfully acknowledge the computer resources, technical expertise, and support provided by the Laboratorio Nacional de Superc\'omputo del Sureste de M\'exico. J.H.-T. acknowledges a student fellowship from the Consejo Nacional de Ciencia y Tecnolog\'ia (Conacyt, M\'exico). The authors thank R. de Coss-Mart\'inez for his critical reading of the manuscript. 

\end{acknowledgments}

\bibliography{ms}

\end{document}


\title{Supplemental Material for\\ ``Understanding the $sp$-magnetism in substitutional doped graphene"}

\author{J. Hern\'andez-Tecorralco}
\email{juanht@ifuap.buap.mx}
\affiliation{Instituto de F\'isica, Benem\'erita Universidad Aut\'onoma de Puebla, Apartado Postal J-48, 72570, Puebla, Puebla, M\'exico
}%
\affiliation{Departamento de F\'isica Aplicada, Centro de Investigaci\'on y de Estudios Avanzados del IPN, Apartado Postal 73, Cordemex, 97310, M\'erida, Yucat\'an, M\'exico
}

\author{L. Meza-Montes}
\affiliation{Instituto de F\'isica, Benem\'erita Universidad Aut\'onoma de Puebla, Apartado Postal J-48, 72570, Puebla, Puebla, M\'exico
}
  
\author{M. E. Cifuentes-Quintal}
\affiliation{Departamento de F\'isica Aplicada, Centro de Investigaci\'on y de Estudios Avanzados del IPN, Apartado Postal 73, Cordemex, 97310, M\'erida, Yucat\'an, M\'exico
}

\author{R. de Coss}
\email{romeo.decoss@cinvestav.mx}
\affiliation{Departamento de F\'isica Aplicada, Centro de Investigaci\'on y de Estudios Avanzados del IPN, Apartado Postal 73, Cordemex, 97310, M\'erida, Yucat\'an, M\'exico
}
\affiliation{Facultad de Ciencias en F\'isica y Matem\'aticas, Universidad Aut\'onoma de Chiapas, M\'exico.}

\date{\today}

\maketitle

\section{Electronic band structure of substitutional doped graphene with spin-orbit coupling}

\begin{figure}[!ht]\centering
	\includegraphics[scale=1.0]{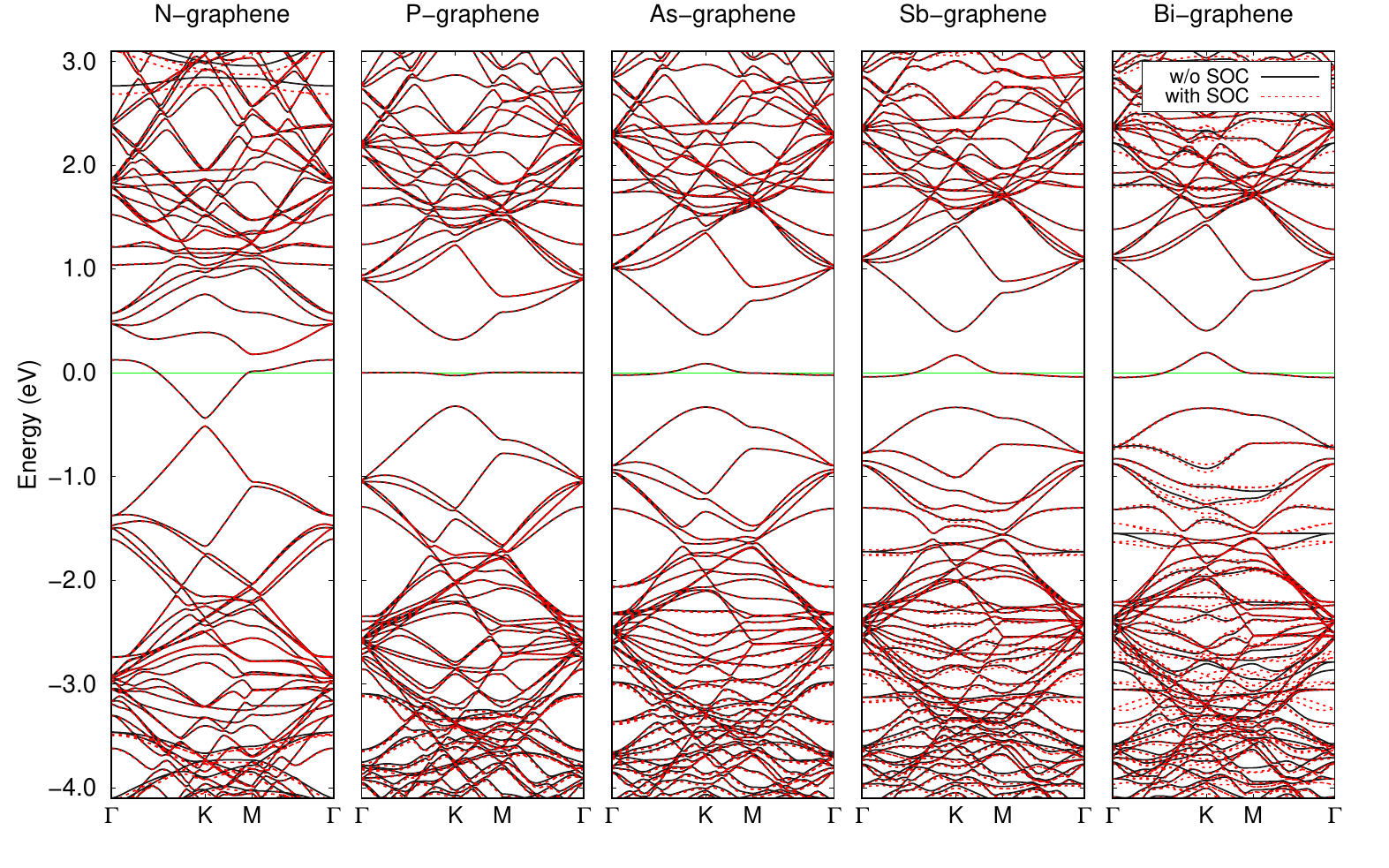}
	\caption{Paramagnetic band structures of doped graphene with \textit{sp}-impurities from VA group (N, P, As, Sb, Bi). The origin of the energy scale has been set at the Fermi level ($E_F$).}
	\label{sup_Fig1}
\end{figure}

The paramagnetic band structures of substitutional doped graphene with N, P, As, Sb, and Bi with and without spin-orbit coupling (SOC) are shown in Figure \ref{sup_Fig1}. The band structures show almost identical behavior around the Fermi level with significant differences only far from it.

 We present the impurity bandwidth dispersion ($W_{imp}$) of the paramagnetic band structure and the total magnetic moment ($M$) obtained from a single self-consistent calculation taking into account the SOC contribution in all the cases. These results are summarized in Table \ref{table1}. 
 
An extra analysis shows that the quantization axis that minimizes the energy is almost degenerate between the $x$, $y$, and $z$ directions. The magnetic anisotropy energy calculation exhibit values smaller than 0.01 meV/supercell. The reported results in table \ref{table1} for the magnetic case corresponds to the quantization axis in the $z$-axis direction. There is no significant difference with the magnetic moment results presented in the main text and the general trends are preserved, considering that they were calculated with the same computational parameters but taking into account SOC and using full-relativistic pseudopotentials \cite{pseudo}. 
 
\begin{table}
\caption{\label{table1} Bandwidth dispersion of the impurity band ($W_{imp}$) and spin magnetic moment ($M$) for each substitutional doped graphene system, calculated including spin-orbit coupling.}
\begin{ruledtabular}
\begin{tabular}{lll}
Impurity & $W_{imp}$ (eV) & $M$ ($\mu_B$/cell) \\
\hline
   N  & 0.563 & 0.00  \\
    P  & 0.032 & 1.00  \\
     As & 0.113 & 0.96  \\
      Sb & 0.209 & 0.66  \\
       Bi & 0.239 & 0.63  \\
\end{tabular}
\end{ruledtabular}
\end{table}
